\documentclass[screen, manuscript]{acmart}

\usepackage{xcolor}
\usepackage{forest}
\usepackage{wrapfig}

\usepackage{listings}
\definecolor{comment-text-color}{rgb}{0,0.8,0.6}

\definecolor{folderbg}{RGB}{124,166,198}
\definecolor{folderborder}{RGB}{110,144,169}

\def\Size{4pt}
\tikzset{
      folder/.pic={
        \filldraw[draw=folderborder,top color=folderbg!50,bottom color=folderbg]
          (-1.05*\Size,0.2\Size+5pt) rectangle ++(.75*\Size,-0.2\Size-5pt);  
        \filldraw[draw=folderborder,top color=folderbg!50,bottom color=folderbg]
          (-1.15*\Size,-\Size) rectangle (1.15*\Size,\Size);
      }
    }

\AtBeginDocument{%
  \providecommand\BibTeX{{%
    \normalfont B\kern-0.5em{\scshape i\kern-0.25em b}\kern-0.8em\TeX}}}

\begin{document}

\title{A backend-agnostic, quantum-classical framework for simulations of chemistry in C\texttt{++}}


\author{Daniel Claudino}
\affiliation{%
  \institution{Computer Science and Mathematics, Oak Ridge National Laboratory}
  \streetaddress{1 Bethel Valley Road}
  \city{Oak Ridge, TN}
  \country{USA}}
\email{claudinodc@ornl.gov}
\thanks{This manuscript has been authored by UT-Battelle, LLC, under Contract No.~DE-AC0500OR22725 with the U.S.~Department of Energy. The United States Government retains and the publisher, by accepting the article for publication, acknowledges that the United States Government retains a non-exclusive, paid-up, irrevocable, world-wide license to publish or reproduce the published form of this manuscript, or allow others to do so, for the United States Government purposes. The Department of Energy will provide public access to these results of federally sponsored research in accordance with the DOE Public Access Plan.}

\author{Alexander J. McCaskey}
\affiliation{%
  \institution{Computer Science and Mathematics, Oak Ridge National Laboratory}
  \streetaddress{1 Bethel Valley Road}
  \city{Oak Ridge, TN}
  \country{USA}}
\email{mccaskeyaj@ornl.gov}

\author{Dmitry I. Lyakh}
\affiliation{%
  \institution{National Center for Computational Sciences, Oak Ridge National Laboratory}
  \streetaddress{1 Bethel Valley Road}
  \city{Oak Ridge, TN}
  \country{USA}}
\email{liakhdi@ornl.gov}


\begin{abstract}
As quantum computing hardware systems continue to advance, the research and development of performant, scalable, and extensible software architectures, languages, models, and compilers is equally as important in order to bring this novel coprocessing capability to a diverse group of domain computational scientists. For the field of quantum chemistry, applications and frameworks exists for modeling and simulation tasks that scale on heterogeneous classical architectures, and we envision the need for similar frameworks on heterogeneous quantum-classical platforms. Here we present the XACC system-level quantum computing framework as a platform for prototyping, developing, and deploying quantum-classical software that specifically targets chemistry applications. We review the fundamental design features in XACC, with special attention to its extensibility and modularity for key quantum programming workflow interfaces, and provide an overview of the interfaces most relevant to simulations of chemistry. A series of examples demonstrating some of the state-of-the-art chemistry algorithms currently implemented in XACC are presented, while also illustrating the various APIs that would enable the community to extend, modify, and devise new algorithms and applications in the realm of chemistry. 

\end{abstract}

\maketitle

\section{Introduction}
Recent experimental results leveraging noisy intermediate-scale quantum (NISQ) computers have demonstrated use cases whereby the quantum processor surpasses conventional computing technologies \cite{google_supremacy, advantage_photonic}, and have provided reassurance for the potential of quantum computing, even though the notion of supremacy/advantage is believed to be intimately tied to tailored operation toward a given domain of application. At the heart of such use cases is a heterogeneous compute model heavily reliant on the interplay between classical computers and quantum co-processors, ultimately implying the necessity for a robust software infrastructure capable of realizing the demanded level of interoperability. Moreover, compliance with vendor-specific native instruction sets and offering an environment that enables hybrid algorithmic development --- all the while remaining highly extensible and easy to maintain --- are some of the key components of a software framework meant to be competitive and outlast the NISQ era into the fault-tolerant regime. These are some of the foundational design features of the XACC framework.~\cite{xacc1}

Modern classical high performance computing has been steadily moving towards an increased level of heterogeneity in both supercomputing centers and clouds. The GPU-accelerated heterogeneous computer architectures are a \textit{de facto} standard of data processing these days. On top of that, more specialized accelerators based on the FPGA and ASIC designs (e.g., Google's TPU) have been developed and already deployed in data centers, addressing specialized workflow needs, for example machine learning and artificial intelligence workloads. To accommodate a variety of computational workloads on diverse hardware, virtualization and service-oriented architectures have become in high performance computing. In the anticipation of even more heterogeneity in future tightly coupled computing platforms enhanced with quantum hardware accelerators, the hybrid quantum-classical programming framework XACC~\cite{xacc1} is designed to serve as a system-level software infrastructure based on a service-oriented architecture. Implemented in modern C++, XACC provides low-latency access to any level of the quantum software stack with a high degree of customization and adaptation to a given hybrid workload via dynamically loaded user-defined plugins. This draws a sharp contrast with most of other quantum programming frameworks which focus on a higher level of language abstraction at the cost of native performance. The XACC framework also provides higher level functionality via both C++ and Python bindings, but it is built on a solid system-level core. Finally, in contrast to many vendor-developed frameworks, XACC is vendor-agnostic, thereby enabling execution of quantum-classical workloads on any physical or virtual backend.

The inherent disposition of quantum bits (qubits) to encode entanglement make natural their usage for simulation of entangled quantum entities, with chemistry being among the most championed applications. There is a wide array of tangible technical areas that are to greatly benefit from an exponential improvement in the speed and scale which computational chemistry can provide a decisive answer, such as drug design,\cite{drug_design1, drug_design2, drug_design3} green catalysts,\cite{catalysis1, catalysis2} and solar cells.\cite{solar_cell1, solar_cell2} While time evolution of a general Hamiltonian is still reserved for devices with a much lower fault rate, the electronic structure of molecules is believed to be addressable within the NISQ paradigm. Such an advance has sparked a great effort in rethinking established formalisms and introducing novel ideas, resulting in new algorithmic strategies for detailing and predicting the electronic structure of systems with chemical interest while exhibiting resource demands that are within the constraints observed in NISQ devices. Advanced software engineering stands to play a critical role in the realization of such ideas, providing the common ground mediating algorithmic development and hardware execution.

With this in mind, in this work we show the utility of XACC as backend-agnostic, hybrid quantum-classical framework that enables development, execution, and benchmarking of chemistry simulations. In the following, we review the core concept of a service-oriented architecture in XACC (Subsection \ref{sec:service}) as the foundational design feature in enabling modularity. This gives the necessary context for a proper presentation of the most relevant interfaces involved in applications of chemical relevance (Subsection \ref{sec:interfaces}). Demonstrations of a variety of quantum algorithms for chemistry simulations are presented and discussed as well as a brief investigation highlighting the superior performance of XACC and its intermediate representation. We close with a summary of the main elements considered throughout the manuscript, along with a brief outlook of how XACC fits into the present and future of quantum software for chemistry applications. 

\section{Service-oriented Architecture}
\label{sec:service}

\begin{figure}[b!]
    \includegraphics[trim= 25 40 0 0,clip,width=\textwidth]{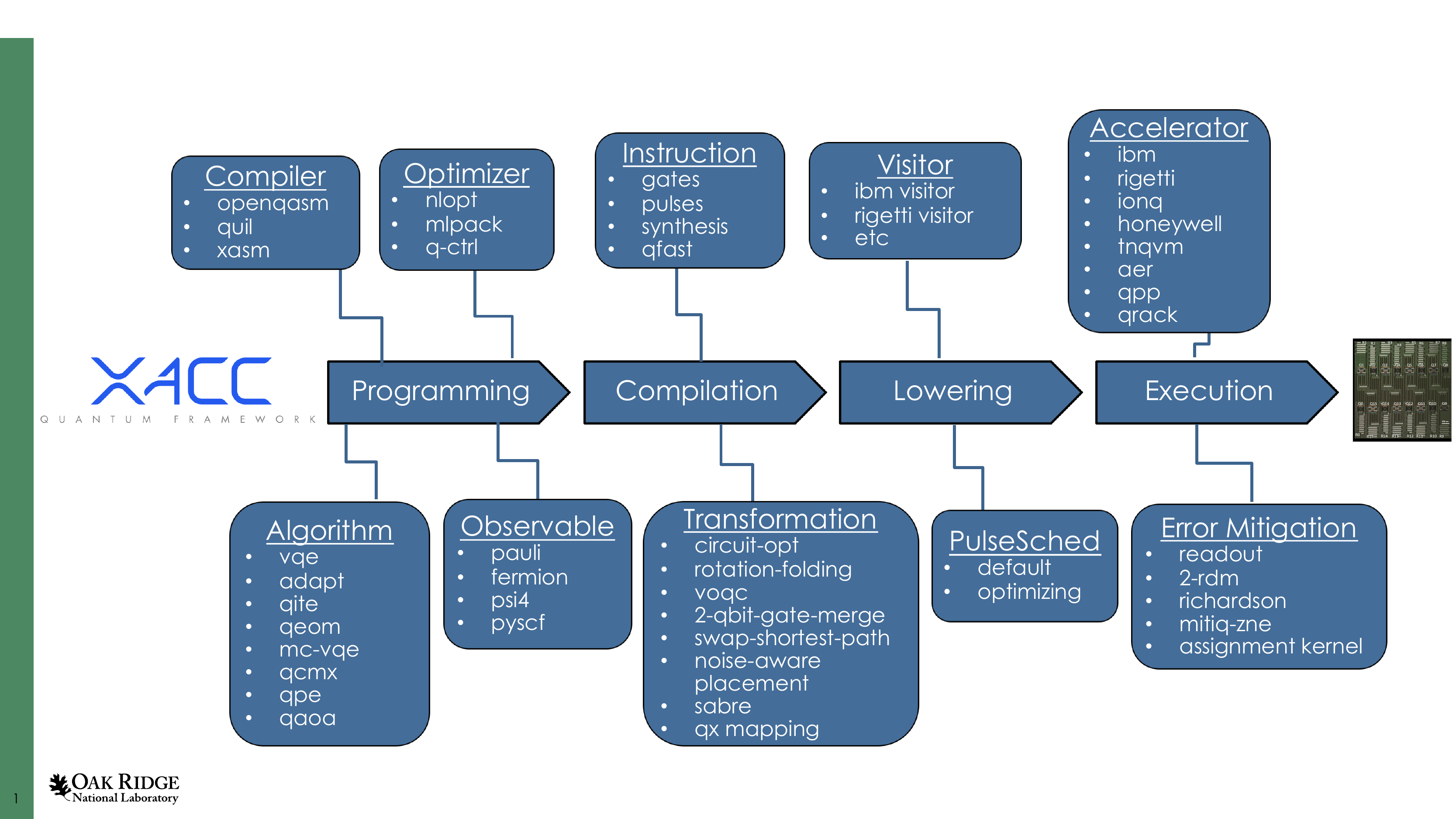}
    \caption{Some pertinent XACC interfaces and concrete implementations at different stages of the quantum computing workflow.}
    \label{fig:workflow}
\end{figure}
Ultimately, XACC puts forward a service-oriented architecture --- the underlying software architecture of XACC defines a collection of core C\texttt{++} classes with pure virtual methods (we henceforth call them \emph{interfaces} or \emph{services} since interface in C\texttt{++} can be modeled via classes with pure-virtual methods) that are intended to be implemented by concrete sub-types and contributed as plugins. XACC puts forward interfaces for language parsing / compiling, describing a domain core intermediate representation, transforming instances of that IR, backend execution, and high-level algorithm description. These interfaces are implemented, compiled into shared-libraries, and contributed to the underlying XACC framework via installation to a pre-specified \emph{plugin} directory. At startup, XACC scans this directory and loads all available plugins, and a corresponding public API call enables programmers to retrieve specific plugin instances corresponding to their unique name and interface type. XACC follows this pattern in an effort to promote maximal extensibility and modularity --- this enables XACC installations to grow over time and adapt or evolve to the rapidly maturing quantum-classical programming, compilation, and execution research field. It also enables developer productivity in that the framework provides core functionality and extensions are provided as simple sub-type definitions. 

To enable this service-oriented architecture, XACC builds upon the CppMicroServices framework~\cite{cppmicroservices} --- a C\texttt{++} native implementation of the Open Services Gateway Initiative (OSGi).\cite{osgi} The OSGi specification has proved extremely successful in the development service-oriented architectures in Java (e.g. the Eclipse IDE), and promotes the concept of a \emph{bundle} as the core unit of extensibility. Bundles declare the services that they provide, and can be deployed in a stand-alone fashion, with the underlying framework implementation understanding how to search for and load available bundle plugins. In XACC, all bundles are deployed as shared libraries containing specific core XACC interface sub-type implementations and code that declares and registers this sub-type with the underlying framework (an \texttt{Activator}). Additionally, all services must inherit from the \texttt{xacc::Identifiable} interface, which requires implementation of a \texttt{name()} method returning the unique name of the plugin sub-type. The core of XACC provides a \texttt{ServiceRegistry} data type that is created at startup and initializes and loads all available plugins and registers them with the underlying CppMicroServices framework instance. The \texttt{ServiceRegistry} exposes a service retrieval API that enables programmers to request plugin instances by name and XACC service interface type (e.g. \texttt{getService<Accelerator>("ibm")}). Through this approach, at the high-level, algorithm programmatic execution consists of the retrieval of pertinent service instances specific to the algorithm, language, and quantum backend desired for the current context.

We now proceed with a brief discussion on the interfaces that are most relevant in chemistry applications.

\section{XACC Interfaces}
\label{sec:interfaces}

XACC enables an extensible workflow for the tasks associated with quantum programming, compilation, and execution via the service-oriented architecture described above. Here we detail the interfaces promoted by XACC that are pertinent for the high-level description and execution of common workflows for quantum chemistry utilizing quantum coprocessors. We note that in this work we seek to provide high-level details on these pertinent interface definitions, but leave a more extensive discussion to \cite{xacc2}.

\subsection{Accelerator and AcceleratorBuffer}
To enable extensibility for backend quantum execution XACC puts forward the \texttt{Accelerator} interface. This interface enables one to inject custom backend implementations for available quantum hardware (IBM, Rigetti, Honeywell, etc.) as well as numerical simulators that scale from laptops to Summit. Ultimately, \texttt{Accelerators} provide an API for one-time backend intitialization and execution of compiled quantum programs. 
The \texttt{initialize()} method takes a map of key-value pairs --- where keys are strings and values can be any type (\texttt{HeterogenousMap}, models a Python \texttt{dict}) --- with the desired parameters to control the accelerator execution, such as number of shots and specific backend type (\texttt{ibmq\_rome} for example), among others. 
The execution of a circuit instance, or a collection of instances, is carried out via \texttt{Accelerator::execute()}, which takes as arguments a qubit register (or an abstraction thereof, see below discussion on \texttt{AcceleratorBuffer}) and the compiled quantum program (see section on XACC IR). The goal of \texttt{execute()} is to map the incoming quantum program to the native instruction set of the targeted backend in the appropriate data format, affect execution of that representation, and post execution results to the provided qubit register buffer. 

XACC currently provides public support for superconducting quantum architectures from IBM and Rigetti and the D-Wave quantum annealers, given the users have the required credentials. Due to difficulties posed by operation of quantum computers in the NISQ era, the employment of virtual simulators is quite common in quantum algorithms research and applications. We offer plugins to several numerical simulators that ship with the XACC installation, among which are the Quantum++ (QPP)\cite{qpp} and Qrack C++ libraries, the Qsim and Aer statevector simulators, and the D-Wave Neal annealer simulator. The TNQVM\cite{tnqvm} plugin exploits the quantum circuit topology and maps quantum circuits onto tensor networks, with a subsequent interface to highly efficient libraries for tensor network manipulations. It also offers the capabilities for density matrix simulations, thereby enabling inclusion of noise channels and backend specific noise models.

The quantum register buffer is modeled through a data type called the \texttt{AcceleratorBuffer}. This type abstracts a register of qubits, with each qubit assigned a unique integer index. Programmers create instances of this data type via a public \texttt{qalloc} call providing the size of the register (number of qubits). Programmers keep reference to this buffer handle and pass it to invocations of \texttt{Accelerator::execute()}, which subsequently persists execution results to the buffer, thereby giving the programmer access these results through the existing buffer handle reference. Another primary role performed by this data type is the storage of execution-specific metadata such as backend noise information, compute statistical values, etc. \texttt{Accelerator} implementations are free to persist any execution-relevant metadata that is then readily available to the programmer with reference to the buffer. 


\subsection{The XACC Intermediate Representation}
XACC provides a polymorphic, extensible intermediate representation for the in-memory expression of parsed and compiled quantum programs. At its core, XACC defines an \texttt{Instruction} interface, which is meant to be sub-typed for concrete quantum instructions. Each \texttt{Instruction} exposes a unique name, the qubit indices that it operates on, and any instruction parameters. Instruction parameters may be provided as string variables or as concrete numeric values, which is useful for the expression of parameterized quantum circuits and templated parameterized quantum circuits (string variable gate rotations which can be made concrete via the evaluation of the variables at numerical values). Moving up the abstraction hierarchy, XACC puts forward the \texttt{CompositeInstruction}, which subtypes \texttt{Instruction} but contains further instructions, enabling a n-ary tree pattern on \texttt{Instruction} types (here nodes in the tree are \texttt{CompositeInstructions} and leaves are concrete \texttt{Instruction} subtypes. On top of these core IR interfaces, XACC also exposes a factory pattern (the \texttt{IRProvider} for the creation of concrete instructions, which ultimately delegates to the core \texttt{ServiceRegistry}. 



\subsection{Optimizer}
\label{ssec:optimizer}

Many use cases of quantum computing in the NISQ era rely on finding a set of parameters that locates an extremum of an objective function. This is foundational in the algorithms grounded on the variational quantum eigensolver, and is also a crucial ingredient in carrying out computation at the pulse level. The problem at hand is defined in terms of an objective function with the general form $F(x_1, \dots, x_n), \quad F: \mathbb{R}^n \rightarrow \mathbb{R}$, such that the quantum processor is responsible for computing $F$ for a certain set of parameters, and $F$ is passed as the argument to a classical optimizer, which in turn checks for convergence and updates the parameters if necessary. To this end, XACC puts forward an \texttt{Optimizer} interface that exposes an \texttt{optimize()} method which takes as input a C++ function with signature \texttt{std::vector<double> x, std::vector<double>\& gradx}. The \texttt{gradx} argument is only used in the event the optimization algorithm conducts parameter search guided by the gradients.

\subsection{Observables, Algorithms, and Gradient Strategies}
In order to describe high-level, domain specific algorithmic expressions we must provide data types enabling common quantum operator or observable creation and manipulation, as well as a unique API for initialization and execution of quantum-classical algorithms. First, XACC exposes the \texttt{Observable} interface, which is meant to provide an abstraction for quantum mechanical observable quantities. Ultimately, one usually expresses these observables in some basis and constructs new instances algebraically. XACC sub-types this interface for pauli or spin-based observable quantities (sums of pauli-tensor product terms) as well as fermionic second quantized representations. At the high level, \texttt{Observables} enable the observation of un-measured quantum circuits. By this we mean that \texttt{Observable} exposes an \texttt{observe} method that takes as input a \texttt{CompositeInstruction} without any concrete \texttt{Measure} \texttt{Instructions} and returns a vector of measured instances of the given \texttt{CompositeInstruction} dependent on the structure of the \texttt{Observable}. In this way, \texttt{Observables} with terms that involve an $\sigma_x$ on a given qubit will result in a new \texttt{CompositeInstruction} with a $\sigma_x$ basis measurement (a Hadamard followed by a Measure-Z). This functionality is leveraged heavily in variational quantum algorithms whereby we have some parameterized quantum circuit and we wish to execute it given some concrete parameters for each term in the observable of interest.
Because fermionic modes are not naturally implemented by quantum hardware, the necessary transformation is made available by the \texttt{ObservableTransform} class, which exposes the \texttt{transform()} method, an example of which is the Jordan-Wigner mapping.\cite{JW} Of note, \texttt{Observables} such as electronic Hamiltonians can be seamlessly generated by interfacing with quantum chemistry packages, where the specification for the quantum chemistry calculation, e.g., basis set, molecular geometry, active spaces, etc., are provided via the \texttt{HeterogeneousMap} with the corresponding keys. XACC currently offers interfaces with the Psi4\cite{psi4} and PySCF\cite{pyscf} suites.

\begin{wrapfigure}{R}{0.5\textwidth}
  \begin{center}
     \includegraphics[width=.5\textwidth]{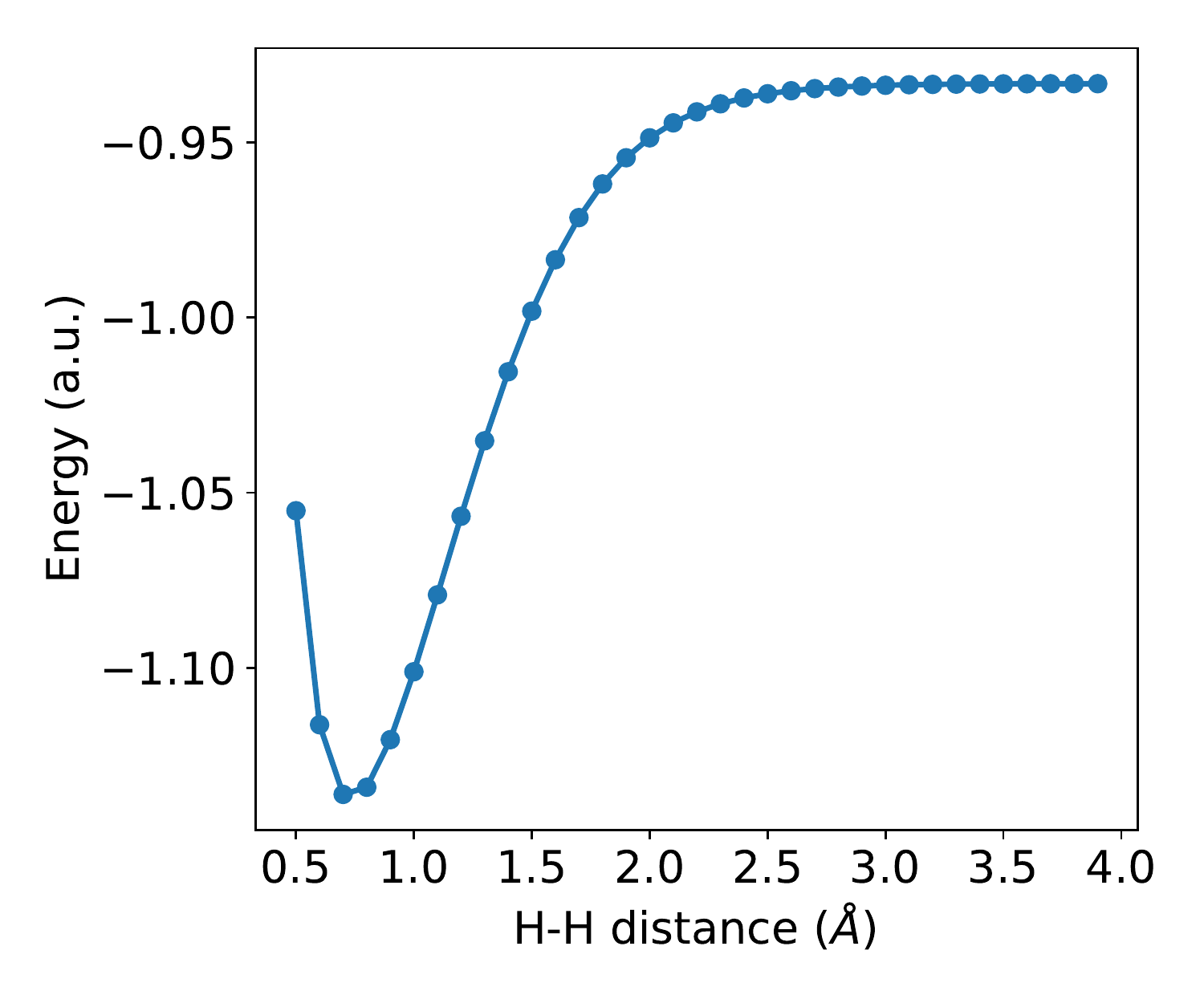}
  \end{center}
\caption{Potential energy curve for H$_2$ in the STO-3G basis set from VQE simulations with the first-order trotter UCCSD ansatz obtained from the program in Figure \ref{snippet:vqe}.}
    \label{fig:vqe}
\end{wrapfigure}
XACC provides an abstraction for algorithm expression, creation, and execution that enables one to contribute and inject new hybrid quantum-classical algorithms as plugin extensions to the underlying framework. The \texttt{Algorithm} interface exposes a mechanism for one-time initialization with pertinent algorithmic input parameters as well as a method for execution of the entire algorithm workflow. XACC exposes \texttt{Algorithm::initialize()} which takes as input a heterogeneous map of key-value pairs that seed the algorithm implementation with pertinent input parameters. For example, this may be the backend to target, the observable to compute expectation values for, and a parameterized \texttt{CompositeInstruction} detailing the ansatz circuit for a variational algorithm. Once initialized, programmers can affect execution of the quantum-classical algorithm via the \texttt{Algorithm::execute()} method, which takes as input an \texttt{AcceleratorBuffer} upon which execution results are persisted. 

As mentioned in Subsection \ref{ssec:optimizer}, XACC offers the possibility of gradient-based optimization. In practical terms, this involves the execution of additional circuits since gradients are computed from varying the rotation angles in the parameterized gates. As such, these circuits can be obtained by calling \texttt{getGradientExecutions()} declared by the \texttt{AlgorithmGradientStrategy} interface whose arguments are the circuit instance (\texttt{CompositeInstruction}) and the current set of parameters $x$. Once these instructions are executed, the derivative with respect to each $x$ is obtained by the \texttt{compute()} method. Currently, XACC enables gradient computation using numerical finite differences (backward, forward, and central differences), parameter shift,\cite{parameter_shift1, parameter_shift2} quantum natural gradient,\cite{qng} and automatic differentiation via the \texttt{AutoDiff} library.
\section{XACC Algorithms for Quantum Chemistry}
\label{sec:demos}
\subsection{Variational Quantum Eigensolver}
\label{ssec:vqe}
The variational quantum eigensolver (VQE)\cite{vqe} has risen as the quintessential example of hybrid quantum-classical algorithm. The preparation and manipulation of highly entangled states, for instance, those that describe interacting fermions, such as in the electronic structure of molecules, pose a level of complexity that classical machines 
\begin{wrapfigure}{r}{0.51\textwidth}
\lstset{language=C++}
\begin{lstlisting}[numbers=right]
#include "xacc.hpp"
#include "xacc_service.hpp"
#include "xacc_observable.hpp"
using namespace xacc::quantum;
int main(int argc, char **argv) {
  // Initialize and load python plugins
  xacc::Initialize(argc, argv);
  xacc::external::load_external_language_plugins();
  
  // instantiate accelerator and optimizer
  auto accelerator = xacc::getAccelerator("qpp");
  auto optimizer = xacc::getOptimizer("nlopt", 
                      {{"algorithm", "l-bfgs"}});

  // instantiate UCCSD ansatz
  auto uccsd = xacc::createComposite("uccsd", 
                        {{"ne", 2}, {"nq", 4}});

  // loop over H-H distances
  for (double r = 0.5; r <= 4.0; r += 0.1) {
    // create hamiltonian
    auto geom = std::string("H 0.0 0.0 0.0\n"
                "H 0.0 0.0 " + std::to_string(r));
    auto hamiltonian = getObservable(
        "pyscf", {{"basis", "sto-3g"}, 
                  {"geometry", geom}});

    // instantiate and initialize VQE
    auto vqe = xacc::getAlgorithm("vqe", 
        {{"ansatz", uccsd},
        {"optimizer", optimizer},
        {"observable", hamiltonian},
        {"accelerator", accelerator}
        {"gradient_strategy", "parameter-shift"}});

    // allocate buffer and execute
    auto buffer = xacc::qalloc(4);
    vqe->execute(buffer);

    // retrieve energy from buffer and print
    auto energy = (*buffer)["opt-val"].as<double>()
    std::cout << energy << "\n";
  }
  
  xacc::external::unload_external_language_plugins();
  xacc::Finalize();
  return 0;
}
\end{lstlisting}
\caption{\label{snippet:vqe}C++ program to compute the potential energy curve of H$_2$ in the STO-3G basis using the VQE algorithm with the first-order trotterized UCCSD ansatz.}
\end{wrapfigure}
have difficulty coping with. In practical terms, VQE casts the variational principle by
targeting the ground state energy from a given Hamiltonian $\hat{H}$ upon preparing a trial state $|\Psi(\vec{\theta}) \rangle $ via a circuit ansatz
with variable parameters $\vec{\theta}$ in the form of single-qubit rotations, and these parameters are varied until optimal rotation angles are found.

This naturally translates into a concerted operation of quantum and classical computers, with the former preparing $|\Psi(\vec{\theta}) \rangle$ and making the required measurements to estimate the value of the objective function, usually the expectation value of the Hamiltonian, for a given set of variational parameters $\vec{\theta}$. The classical computer receives and processes the outcomes of the measurements, which serve as input for an optimization routine that checks whether the parameters that minimize the objective function have been found.

A popular ansatz choice derives from the unitary coupled cluster theory, which is not feasible for direct implementation in quantum hardware and needs to be approximated, often to low-order in the Trotter-Suzuki approximation (also known as Trotterization).\cite{Hatano2005} We illustrate the usage of VQE in the case of the potential energy curve of H$_2$ with the STO-3G basis set where we prepare the ground state according to the first-order Trotterized UCCSD ansatz in Figure \ref{snippet:vqe}.

In this first example, it is beneficial to highlight and explain many of the APIs that will be recurrent in the demonstrations to follow. In lines 1-3 we include the necessary headers: \texttt{xacc\_service.hpp} provides the \texttt{getService<typename Interface>(serviceName)} API to load service plugins and \texttt{xacc\_observable.hpp} allowing creation of a general object of the \texttt{Observable} type. The former originates from the interface with PySCF, a Python native package, which are loaded and unloaded as shown in lines 8 and 50, respectively. Getting a reference to many of the most common interface types is facilitated by wrappers of the sort \texttt{xacc::getInterface(serviceName, options)}, which are illustrated in lines 11 (\texttt{Accelerator}), 12-13 (\texttt{Optimizer}), 24-26 (\texttt{Observable}), and 29-34 (\texttt{Algorithm}), where \texttt{options} is a \texttt{HeterogenousMap} with the parameters specific to the service in question. In this particular example, we use the virtual backend accelerator building on the \texttt{Quantum++}\cite{qpp} library (line 11), the L-BFGS optimization algorithm in NLOpt (line 14), and the first-order trotterized UCCSD ansatz, whose instantiation is shown in lines 18-19, with the keys ``ne'' and ``nq'' taking as values the number of electrons and number of qubits, respectively. Within the \texttt{for} loop enclosed in lines 22-47, the geometry of the H$_2$ molecule and the corresponding \texttt{Observable} is updated in lines 25-29. With all the necessary parameters, we get a reference to the VQE algorithm, initialized with all the previously instantiated parameters as well as a \texttt{string} that identifies the desired gradient strategy, which in this case is parameter shift. 

The energy is persisted in the buffer as a \texttt{double} and can be retrieved with the proper key (``opt-val'') and knowledge of the associated type. The resulting potential energy curve is plotted in Figure \ref{fig:vqe}.

\subsection{ADAPT-VQE}

\begin{figure}[h!]
\minipage{0.48\columnwidth}
\lstset{language=C++}
\begin{lstlisting}[numbers=left]
// instantiate accelerator, optimizer
// and allocate buffer

// instantiate observable in an active space
std::vector<int> frozen = {0,6}, 
    active = {1, 2, 3, 4, 5, 7, 8, 9, 10, 11};
auto hamiltonian = 
    xacc::quantum::getObservable("pyscf", 
    {{"basis", "sto-6g"},
    {"geometry", geom},
    {"frozen-spin-orbitals", frozen},
    {"active-spin-orbitals", active}});

// instantiate and initialize ADAPT-VQE
auto adapt = xacc::getAlgorithm("adapt",
                {{"optimizer", optimizer},
                {"observable", hamiltonian},
                {"sub-algorithm", "vqe"},
                {"n-electrons", 4},
                {"pool", "singlet-adapted-uccsd"},
                {"accelerator", accelerator}});

// execute
adapt->execute(buffer);
\end{lstlisting}
\endminipage
\hfill
\minipage{0.48\columnwidth}
\includegraphics[width=\columnwidth]{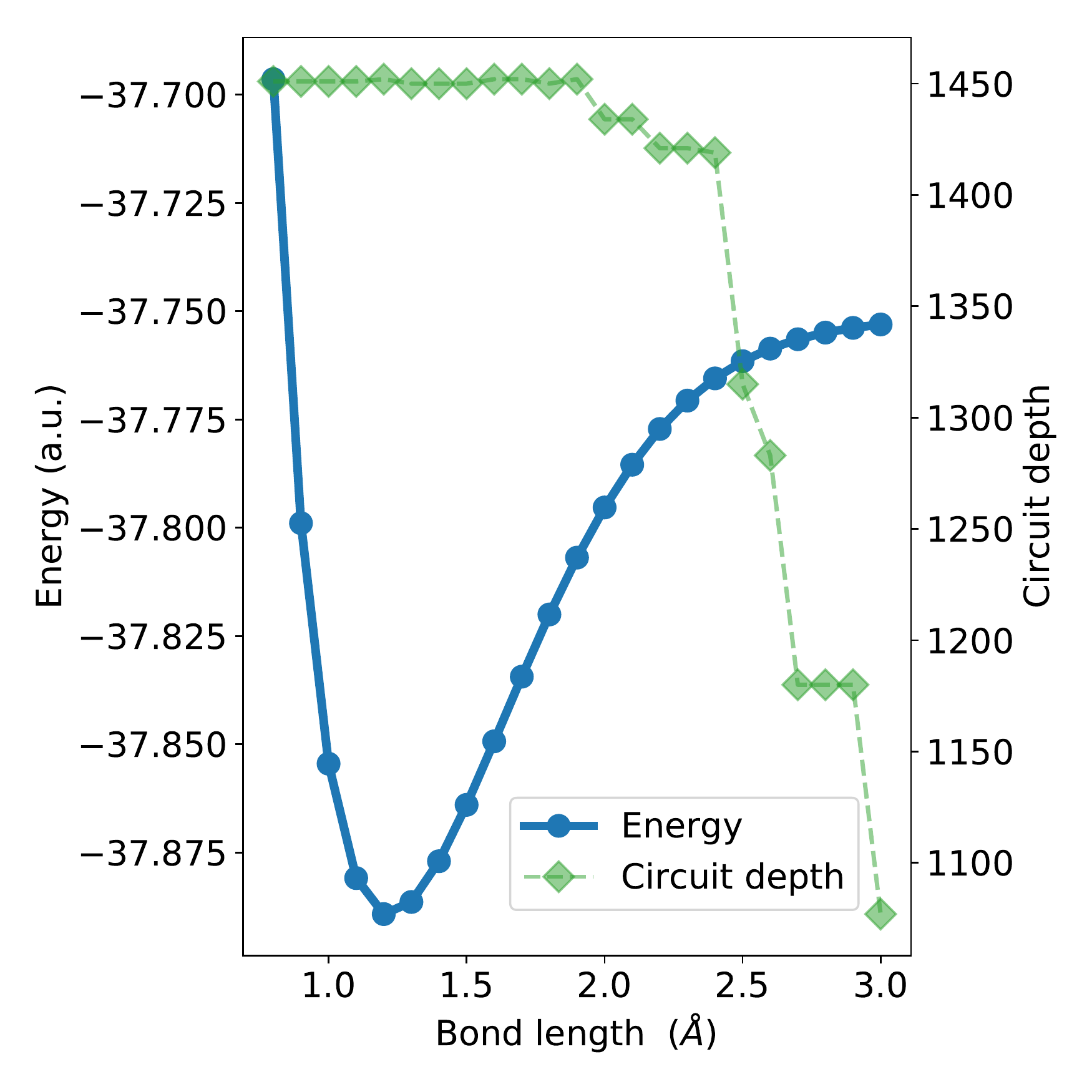}
\endminipage
\caption{\label{snippet:adapt}Code snippet highlighting the ADAPT-VQE code initialization and execution for H$_4$ with the singlet-adapted single and double excitation operators (left) and the potential energy curve of CH$^+$ in the STO-6G with the lowest energy spatial orbital frozen. For comparison, the circuit depth of the first-order trotterized UCCSD ansatz for the same molecule is 11665.}
\end{figure}
Despite being able to execute certain tasks exponentially faster than its classical counterparts, quantum computers are at a stage that there is only a limited time window in which this can be done before the information stored in the qubits is lost due to external effects, collectively referred to as ``noise'' . This time window is known as the coherence time and its narrow width is a major impediment in the maturing of quantum information technologies as a whole. In practical terms, this caps the so-called circuit depth, which correlates to the amount of successive operations that can be performed by a quantum circuit. In an effort to work around this limitation, the Adaptive Derivative-Assembled Pseudo-Trotter (ADAPT)\cite{adapt, qubit_adapt, adapt_qaoa} ansatz replaces the alternatives based on the straightforward application of the UCC by capitalizing on the Trotter-Suzuki formula. An ADAPT-VQE simulation of the potential energy curve of CH$^+$ ion in the singlet electronic state can be set up with the code snippet in Figure \ref{snippet:adapt}, along with the output energy values and circuit depths for different bond lengths.

\begin{wrapfigure}{O}{0.53\columnwidth}
    \lstset{language=C++}
    \begin{lstlisting}[numbers=right]
// Hamiltonian
using namespace xacc::quantum;
auto H = PauliOperator(0.2976);
H += PauliOperator({{0, "Z"}}, 0.3593);
H += PauliOperator({{1, "Z"}}, -0.4826);
H += PauliOperator({{0, "Z"}, {1, "Z"}}, 0.5818);
H += PauliOperator({{0, "X"}, {1, "X"}}, 0.0896);
H += PauliOperator({{0, "Y"}, {1, "Y"}}, 0.0896);
Observable* observable = &H;

// Get reference to the TNQVM Accelerator
auto accelerator = xacc::getAccelerator("tnqvm");

// Map quantum kernel code into XACC IR
// and compile to accelerator
auto compiler = xacc::getCompiler("xasm");
auto ansatz = compiler->compile(R"(
__qpu__ void kernel(qbit q) {
  X(q[0]);
})", accelerator)->getComposite("kernel");

// Time propagation parameters
int nSteps = 100;
int stepSize = 0.05;

// Initialize QITE
auto qite = xacc::getAlgorithm("qite", 
                        {{"accelerator", accelerator}, 
                        {"observable", observable}, 
                        {"step-size", stepSize},
                        {"steps", nSteps}, 
                        {"ansatz", ansatz}});
                            
// Allocate buffer and execute
auto buffer = xacc::qalloc(2);
qite->execute(buffer);    
\end{lstlisting}

\caption{\label{snippet:qite}Code snippet for the QITE simulation of H$_2$ at H-H distance of 0.7{\AA} with the Hamiltonian from Ref.\cite{scalable}.}
\end{wrapfigure}
The core assumption in the ADAPT-VQE algorithm is that an easy-to-prepare state, normally a mean-field approximation such as Hartree-Fock, can be connected to the ground state with degree of accuracy by only a limited number of many-body rotations drawn out of an ``operator pool''. With the example of direct application of the UCCSD ansatz in mind (Subsection \ref{ssec:vqe}), it aims at retaining only the operators that provide a sizable ground state energy improvement, emphasizing double excitations over singles. At each ADAPT iteration, the gradient with respect to the operator pool, which equals the commutators of the Hamiltonian with the operators in the pool, have to be measured, and the norm of this gradient controls the number of iterations the ADAPT algorithm goes through. If this norm is not found below a user-defined threshold, the operator displaying the largest commutator (in absolute terms) is introduced in the ansatz, whose parameters are optimized by a usual VQE call. This can greatly reduce circuit depth by only retaining the many-body rotations required to reach the ground state energy (to some precision). On the other hand, the relative importance of each operator as determined by their commutators needs to be re-evaluated at each iteration, shifting the demand for quantum resources to the measurement task. 

\subsection{Quantum imaginary time evolution}

The two algorithms above highlight one frequent impasse in the choice of a variational quantum algorithm, which is the trade-off between circuit depth and number of measurements. An alternative to methods ground on the variational principle is found in the family of approaches stemming from imaginary time evolution.\cite{McArdle2019, motta2019determining} It has the benefit of targeting eigenstates as well as thermal states on an equal footing in an efficient fashion as long as the time propagation as done slowly or, in other words, the propagated terms remain relatively local. 

\begin{wrapfigure}{R}{0.5\textwidth}
  \begin{center}
     \includegraphics[width=.5\columnwidth]{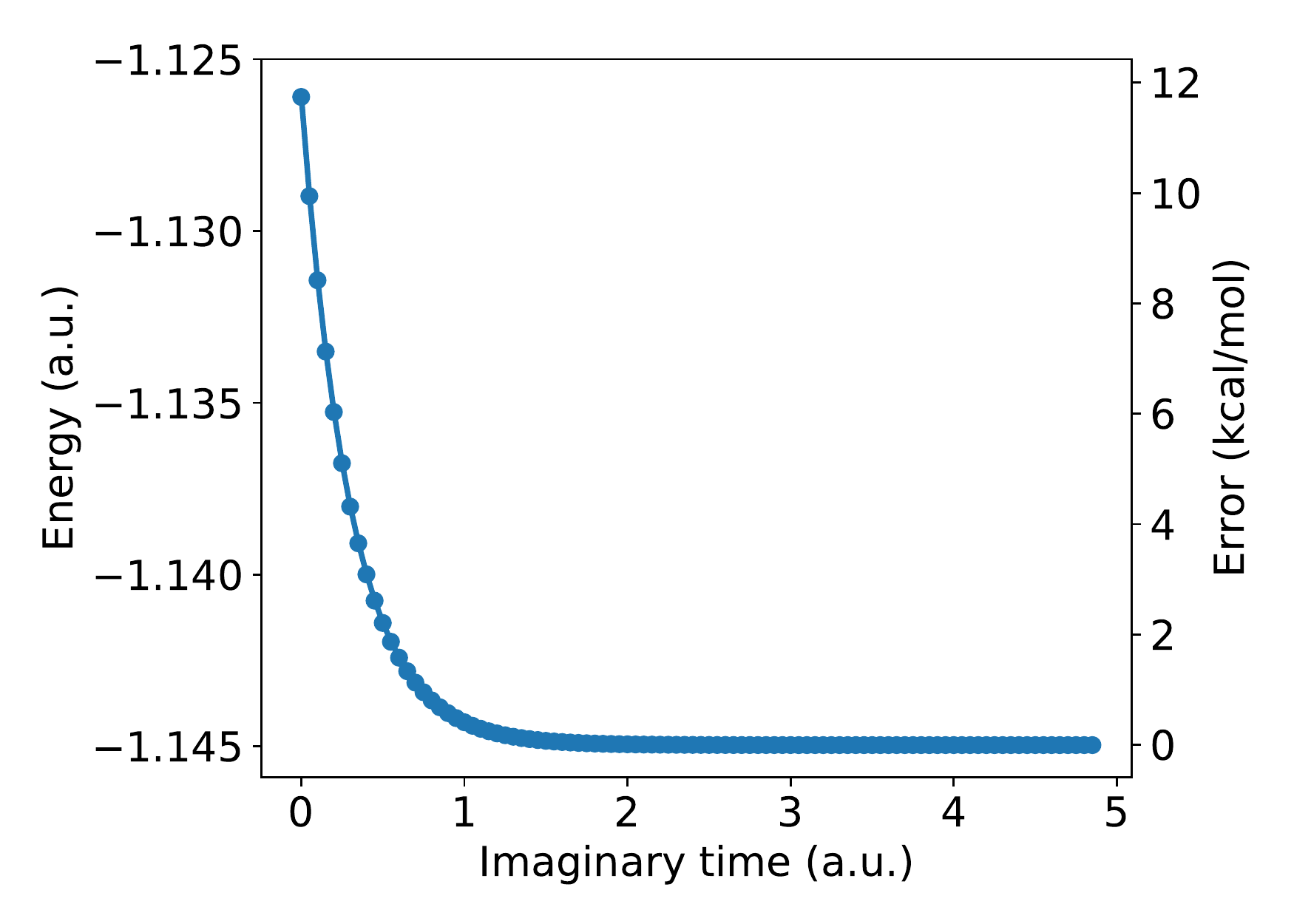}
  \end{center}
  \caption{Ground state energy H$_2$ in the STO-6G basis set throughout imaginary time evolution.}
  \label{fig:qite}
\end{wrapfigure}

The code snippet provided in Figure \ref{snippet:qite} can be used to reproduce the plot in Figure \ref{fig:qite}. This is a simple use case of the QITE algorithm that displays the convergence of the ground state energy throughout propagation of the H$_2$ Hamiltonian in the STO-6G in imaginary time, which is shown in Figure \ref{fig:qite}. We also note that the closely quantum analog of the Lanczos method to target specific eigenvalues is also available in XACC. Here we present yet another way to construct compound observables, showcasing some of the operator algebra underlying the \texttt{PauliOperator} type in lines 1-7 in Figure \ref{snippet:qite}, whose overloaded constructor enables several variants of Pauli operators to be accessed. We also show an alternative way to get the \texttt{CompositeInstruction} object that represents the circuit to be employed in the simulation. Upon creating instances for the accelerator and compiler, a string literal representing the quantum kernel to be compiled is passed to the \texttt{xacc::Compiler::compile()} method, which maps the string into an instance of XACC IR and further process it if necessary before compilation to the accelerator, with the resulting \texttt{CompositeInstruction} being retrieved by calling \texttt{getComposite()} such that the argument matches the name of the quantum kernel (``kernel'' here). In Figure \ref{fig:qite} we can see how the ground state energy of H$_2$ in the STO-6G evolves in imaginary time by propagating the HF state through 100 imaginary time steps of 0.05 units of time each.

\subsection{Quantum moments expansions}

\begin{figure}[h!]
\minipage{0.53\columnwidth}
\lstset{language=C++}
\begin{lstlisting}[numbers=left]
// instantiate accelerator, optimizer, and observable
auto nOrbitals = observable->nBits();

// instantiate ansatz that prepares the HF state
auto irgen = 
    xacc::getService<xacc::IRProvider>("quantum");
auto ansatz = irgen->createComposite("initial-state");
for (auto i = 0; i < nElectrons / 2; i++) {
  auto alphaX = irgen->createInstruction("X", {i});
  ansatz->addInstruction(alphaX);
  auto betaX = irgen->createInstruction("X", 
                            {i + nOrbitals / 2});
  ansatz->addInstruction(betaX);
}

// instantiate and initialize QCMX for some cmx_order
auto qcmx = xacc::getAlgorithm("qcmx", {
                        {"ansatz", ansatz},
                        {"accelerator", accelerator},
                        {"observable", observable},
                        {"cmx-order", cmx_order}
                        });
                        
// allocate buffer and execute
\end{lstlisting}
\endminipage
\hfill
\minipage{0.45\columnwidth}
\includegraphics[width=\columnwidth]{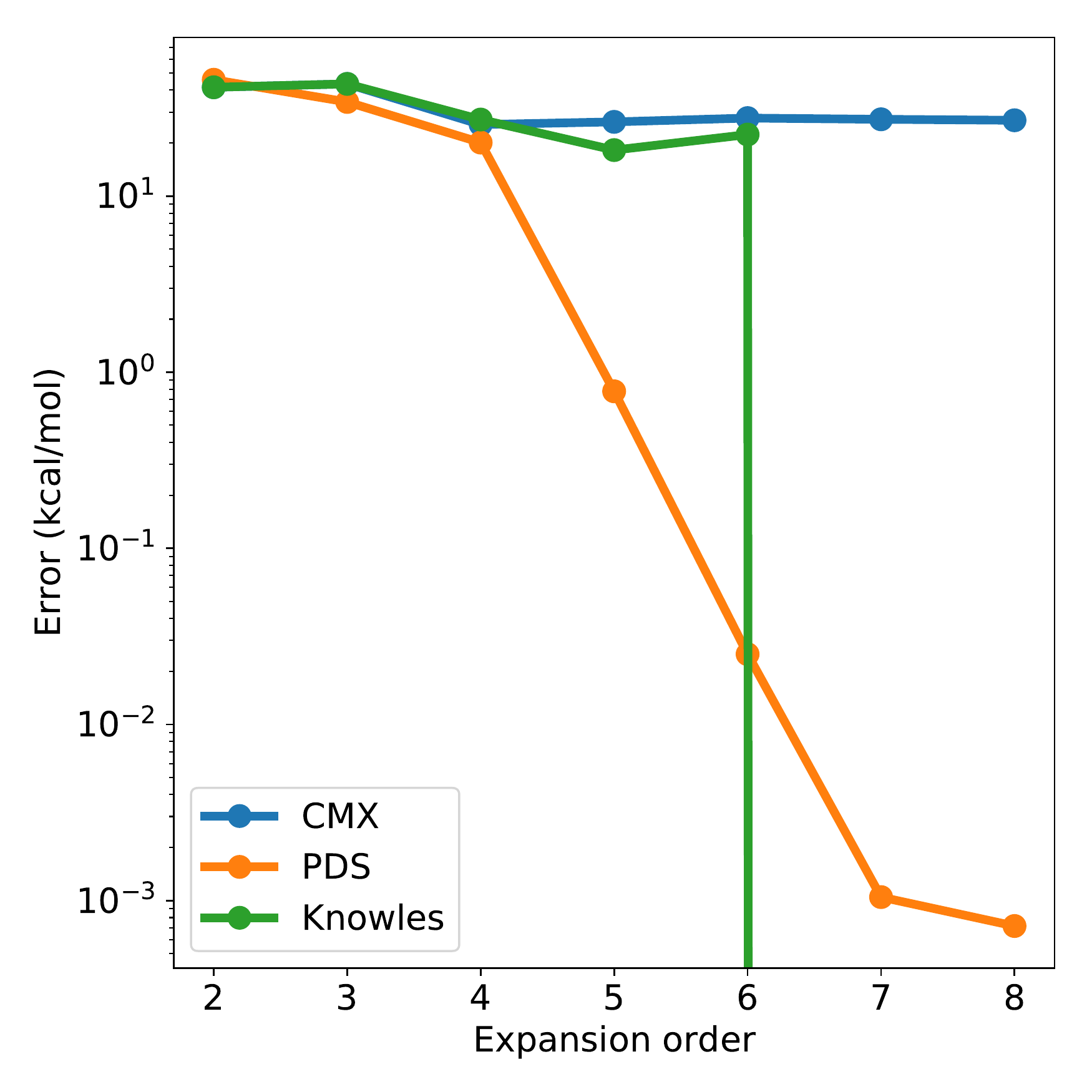}
\endminipage
\caption{\label{qcmx}Code snippet highlighting the ADAPT-VQE code initialization and execution for H$_4$ with the singlet-adapted single and double excitation operators (left) and the potential energy curve of CH$^+$ in the STO-6G with the lowest energy spatial orbital frozen. For comparison, the circuit depth of the first-order trotterized UCCSD ansatz for the same molecule is 11665.}
\end{figure}

Methods grounded on expansions of the moments of the Hamiltonian have a long history that dates back to the ``$t$-expansion'' in terms of connected moments.\cite{horn1984t} While it provides the exact ground state energy in the limit $t\rightarrow \infty$, several approximations have been proposed over the years in order to establish energy estimates for finite $t$.\cite{cioslowski1987connected, knowles1987validity, peeters1984upper, soldatov1995generalized} Several connections between moments expansions, power methods,\cite{seki2020quantum} and imaginary time evolution can be drawn, each also displaying their particularities. At the heart of this family of approaches lies the computation of matrix elements of the type $\langle H^k \rangle$, which can be in turn employed in the computation of connected moments appearing in some of these methodologies. Computation of such quantities can expected to be performed efficiently by quantum computers, thus opening the door to quantum variants of moments expansions.\cite{Kowalski2020, peng2021variational, claudino2021improving} We show in Figure \ref{qcmx} a sample program to perform such simulations.

Considering that we map each spin-orbital in the Hamiltonian to a qubit, the number of orbitals can be simply retrieved as the number of bits in the observable, as shown in line 2 of the snippet in Figure \ref{qcmx}. Taking the square H$_4$ molecule in the STO-6G basis set where the distance between adjacent hydrogen atoms is 2{\AA}, we end up with 8 spin-orbitals. The circuit to prepare the Hartree--Fock (HF) state is obtained following lines 5-12. In XACC we follow the convention that all alpha spin-orbitals are mapped to the first half of the qubits in the register, followed by the beta spin-orbitals, thus resulting in $|11001100\rangle$ as the computational basis state representation of the  HF state. The QCMX algorithm computes the energy of all orders from 2 up to \texttt{cmx\_order} (line 18) using three different expansion types: Cioslowski's connected moments expansion (CMX), the Peeters--Devreese--Soldatov (PDS) energy functional, and the Knowles' generalized Pad\'e approximants, while many others can be implemented. Going up to order 8 for the H$_4$ molecule above, we obtain the errors from the exact ground state energy reported in Figure \ref{qcmx}.

\subsection{Quantum equation of motion}

\begin{wrapfigure}{O}{0.52\textwidth}
\begin{lstlisting}[numbers=right]
// instantiate accelerator, ansatz, and observable
// get number of electrons (nElectrons)
auto qeom = xacc::getAlgorithm("qeom", 
                        {{"ansatz", ansatz}
                        {"accelerator", accelerator},
                        {"observable", observable},
                        {"n-electrons", nElectrons}});
auto q = xacc::qalloc(observable->nBits());
qeom->execute(q);
\end{lstlisting}
    \caption{Code snippet showing the required parameters for the QEOM algorithm.}
    \label{snippet:qeom}
\end{wrapfigure}
A great deal of effort in quantum chemistry revolves around finding suitable estimates for the ground state and its energy. For this reason, VQE has been quite popular for chemistry simulations since its inception due to its foundation on the variational principle. It has been shown it can be used in the search for excited states, yet for this application it requires certain deviations from its original formulation, such as the introduction of penalty terms to prevent the solver to enter the symmetry subspace where the ground state solution can be found.\cite{penalty} Borrowing from the equation-of-motion (EOM) formalism and its classical implementations,\cite{eom1, eom2} the quantum version of the EOM method (QEOM)\cite{qeom} yields the spectrum of electronic excitations and de-excitations solving a secular system equations in the basis of the excitations operators that span the targeted excitation subspace where the desired state manifold can be found. Once the required matrix elements are computed from a provided approximation to the ground state, the energy spectrum follows classically from the solution of a generalized eigenvalue problem. The preparation of the input ground state precedes the QEOM algorithm and is largely detached from it, so there is great freedom in choosing how to approximate the ground state, e.g., VQE.

\begin{figure}
    \centering
    \includegraphics[width=.75\columnwidth]{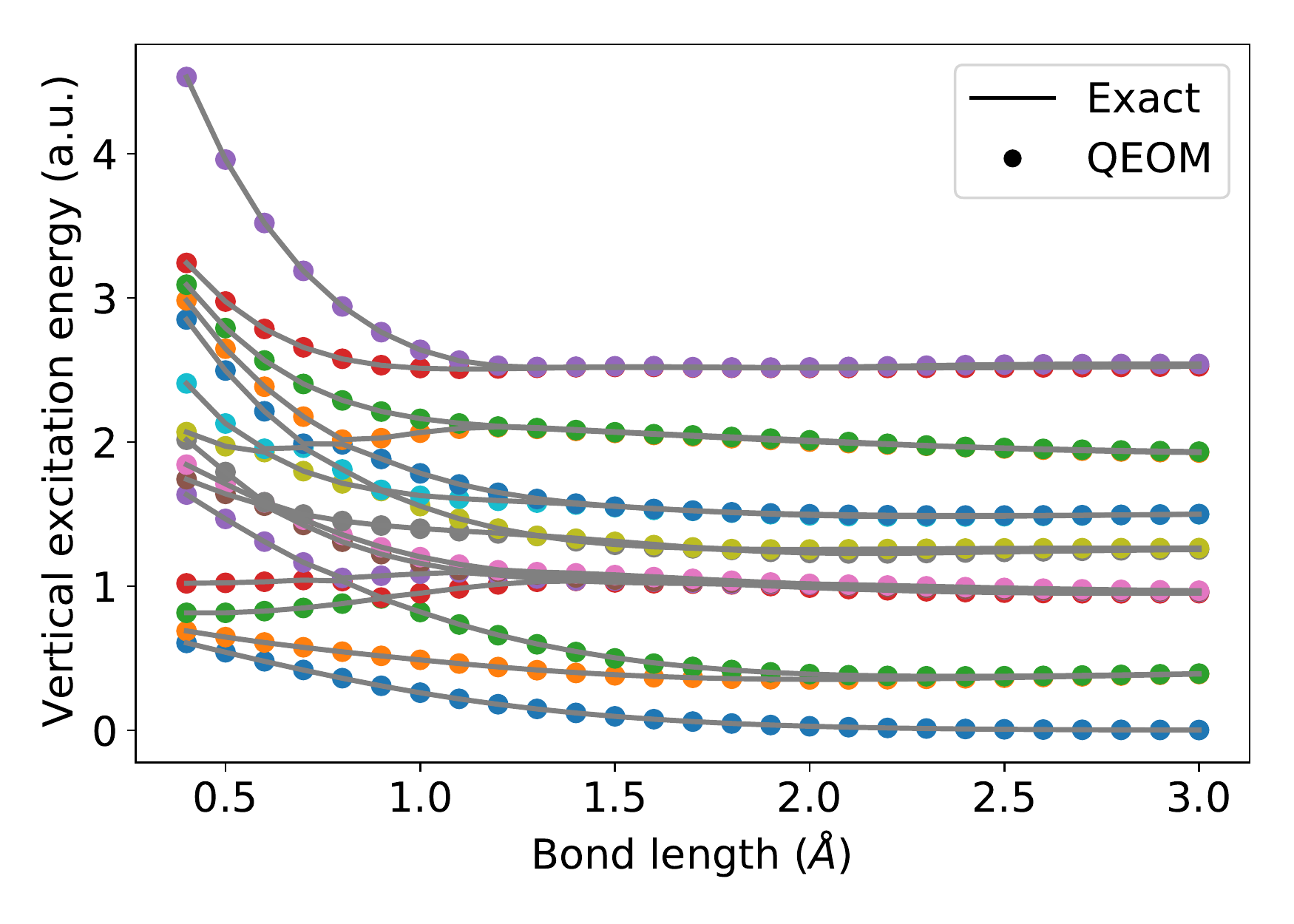}
    \caption{Potential energy curves for the electronic excited states of the H$_2$ molecule in the 6-31G basis set (particle-number conserving subspace).}
    \label{fig:qeom}
\end{figure}

An example showing the parameters that are required by the QEOM algorithm is displayed in Figure \ref{snippet:qeom}. Preparing the approximate ground state of the H$_2$ molecule in the 6-31G basis set with 4 ADAPT-VQE iterations and simulating the electronic transitions using the QEOM algorithm results in the potential energy curves plotted in Figure \ref{fig:qeom}.

\subsection{MC-VQE}

Some properties exhibited by electrons make them particularly demanding in their quantum-mechanical treatment, such as the antisymmetry of the wave-function under exchange of indistinguishable fermionic modes and the long-range Coulomb interaction between any two electrons. However, in the limit where these electrons are fairly constrained to disjoint spatial regions, these requirements can be relaxed to a good approximation. This is at the heart of the ab initio exciton model (AIEM)~\cite{aiem1} which offers a possibility of modeling excited states in large molecular systems formed by weakly coupled units, or monomers, for example in biological chromophore complexes. The multi-contracted VQE (MC-VQE)\cite{mc-vqe} quantum algorithm casts the AIEM Hamiltonian into the basis of Pauli operators and, by restricting the number of states in each monomer to their ground and first excited states, enables the collective absorption spectrum to be obtained via two-qubit gates entangling qubits corresponding to isolated monomers. It deviates from the usual VQE due to 1) minimizing the state-averaged energy while all states share the same entanglers, and 2) the final spectrum comes from a classical diagonalization in the so-called ``interference'' state basis. 

\begin{figure}
    \centering
    \includegraphics[width=\columnwidth]{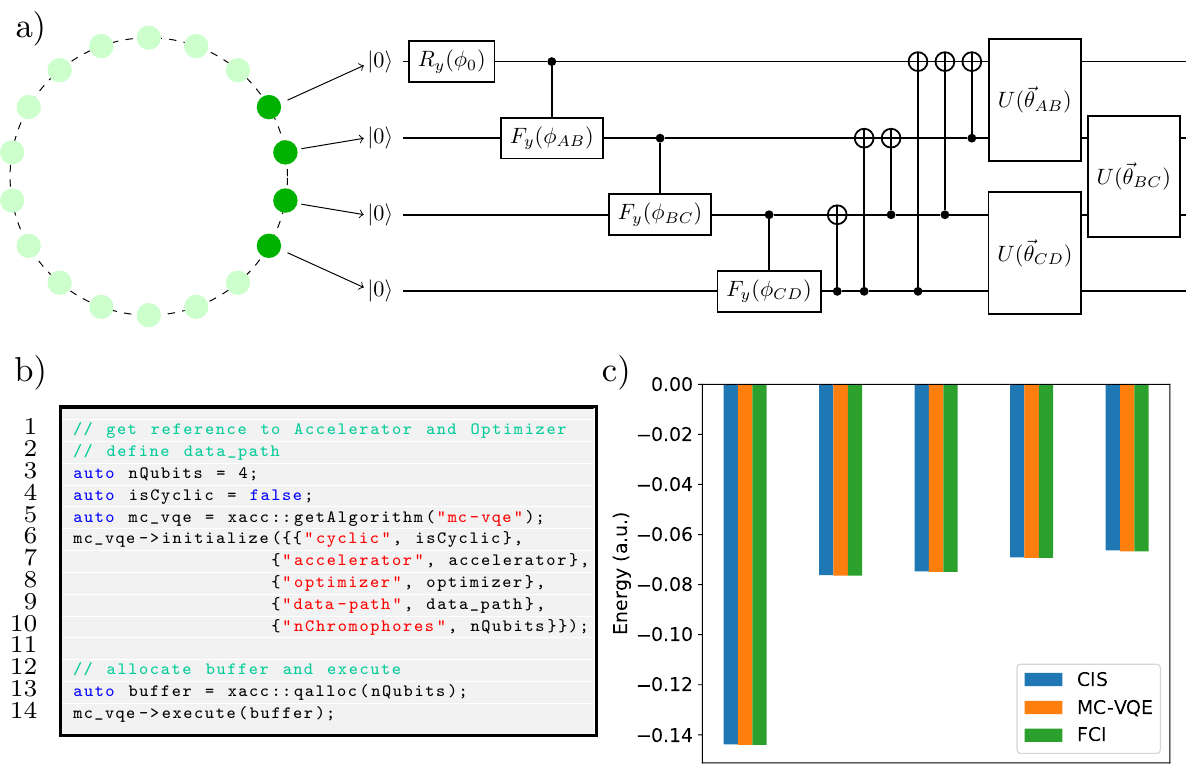}
    \caption{\label{fig:mcvqe}a) Pictorial representation of the LH2 B850 ring complex and the selected segment with the associated MC-VQE state preparation circuit; b) Code snippet illustrating MC-VQE; c) Comparison of the resulting spectrum among configuration interaction singles, MC-VQE, and full configuration interaction.}
\end{figure}

In the MC-VQE algorithm, the ground and first excited state of each chromophore map onto the states of a dedicated qubit, so the number of chromophores (line 3) determines the size of the qubit register. Besides some of the recurrent \texttt{HeterogeneousMap} keys, the MC-VQE takes the \texttt{cyclic} key to inform if the chromophore connectivity is cyclic or linear, which in turn translates into whether or not the qubits indexed by 0 and \texttt{nQubits - 1} should be entangled. The value associated to the \texttt{data-path} is the directory where the classical data (electronic energies, dipole moments, etc.) is located, which are processed with the help of an \texttt{Importable} helper class. For example, we can simulate a 4-chromophore segment of the LH2 B850 ring complex, with a sample multi-contracted state preparation circuit being shown in Figure~\ref{fig:mcvqe}a). Figure~\ref{fig:mcqve}b) highlights the code pertinent to the MC-VQE algorithm (classical quantities needed to compute the AIEM Hamiltonian are obtained from the Supplemental Material in Ref.~\cite{fig:mcvqe}).The resulting absorption spectrum, compared with the configuration interaction singles (CIS) and exact diagonalization (full configuration interaction, FCI) results are displayed in Figure~\ref{fig:mcvqe}c).

\subsection{Comparison with Python-based framework}

\begin{figure}
    \centering
    \includegraphics[width=\columnwidth]{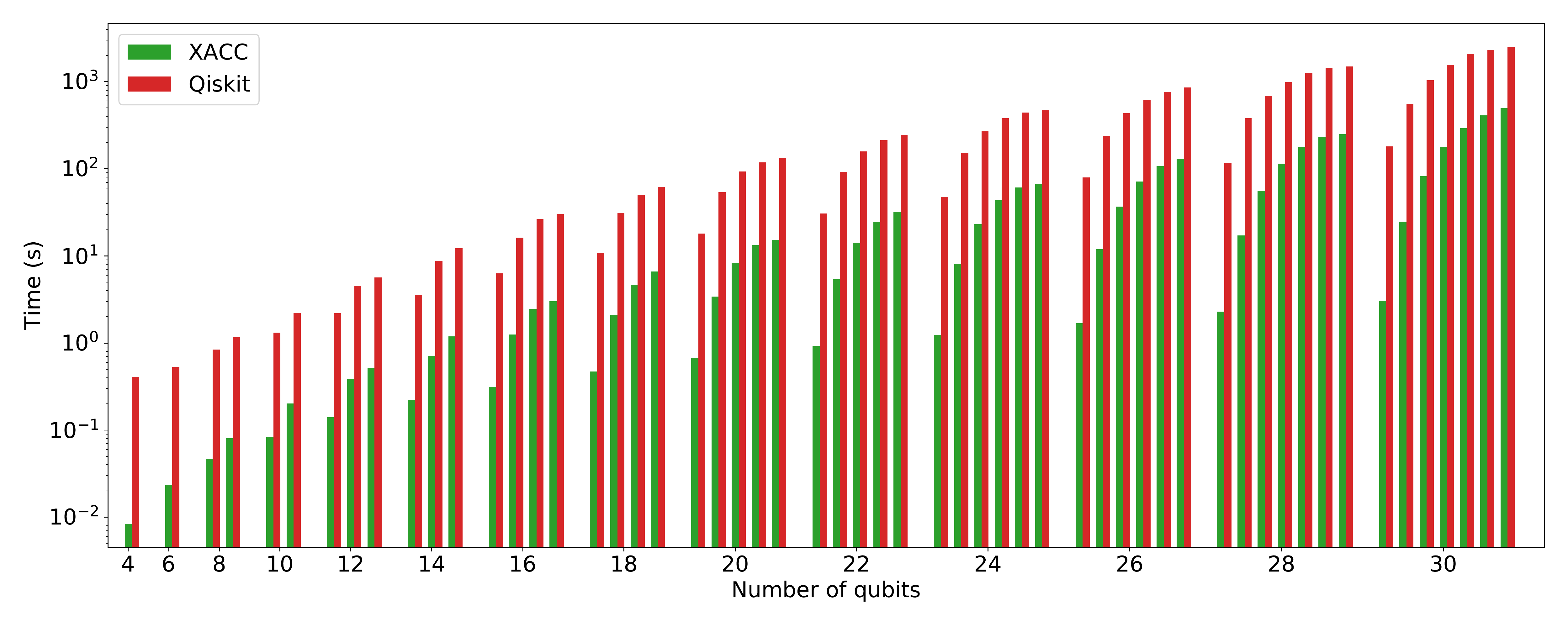}
    \caption{Timings for the construction of the first-order trotterized UCCSD circuit for XACC and Qiskit.}
    \label{fig:qiskit}
\end{figure}

The fact that the vast majority of virtual backends (numerical simulators) are written in low-level languages attests to the necessity of a high degree of resource control that they enable. Yet, most software stacks for chemistry simulations in quantum processors, or quantum computation more generally, are devised in high-level languages, with special focus on Python. Despite the notion that the backend is assinged the most arduous tasks, there are still benefits from adopting more perfomant languages throughout the many stages along the workflow. To corroborate this claim, we take IBM's Qiskit~\cite{Qiskit} to be representative of a Python-based framework and compared it against XACC in that time they take to construct the circuit that prepares the state associated with the first-order trotterized UCCSD ansatz for several combinations of qubit registers and number of electrons. The best timings out of 10 runs are reported in Figure \ref{fig:qiskit}.

The number of operators/gates in the UCCSD ansatz is dominated by double excitations. For $N_q$ the number of qubits and $N_e$ the number of electrons, the total number of double excitations is 
\[{N_q/2\choose N_e}{N_q/2\choose N_e} = {N_q/2\choose N_q/2 - N_e}{N_q/2\choose N_q/2 - N_e}\]
thus we only report the timings for $N_e < N_q/2$ for a given $N_q$ in Figure \ref{fig:qiskit}. Here again we can see additional evidence for harnessing the power of the more robust C\texttt{++} and the XACC IR it enables, since it leads to circuit constructions that are at least an order of magnitude faster than the analogous process with Python as performed by Qiskit.

\section{Summary and Outlook}

We have presented a collection of state-of-the-art algorithms for quantum chemistry leveraging quantum co-processors, as well as their implementation and usage within the XACC framework, thereby allowing users to immediately have access to algorithms covering a wide variety of chemical phenomena. Furthermore, it is important to emphasize that the core functionality of XACC is largely exposed through an extensive list of Python bindings, facilitating utilization by users more familiar with such a language.

Even more importantly, with the preceding discussion of the main XACC interfaces and its comprehensive list of APIs, we hope to have conveyed that XACC stands out for its ability and flexibility in enabling development with the goal of high-performance and potential deployment across multiple quantum hardware platforms and that these features can be exploited in devising, implementing, and benchmarking quantum algorithms for chemistry, all the while fostering an open-source environment. In passing, it is also noteworthy that many other domain sciences with a common ground in quantum mechanics can equally benefit from the infrastructure made available by XACC, such as quantum field theories, solid state physics, quantum dynamics, etc.

The ecosystem of software tools for quantum computation targeting chemical applications has been largely influenced by Pythonic frameworks, with some of the notable examples being Qiskit\cite{Qiskit}, OpenFermion\cite{openfermion}, PennyLane\cite{pennylane}, and TEQUILA\cite{tequila}. With the expected increase of orders of magnitude in the number of qubits envisioned for the next few years and the applications that they will likely enable comes an ever growing need for a robust and lasting software infrastructure, which strengthens the case for system-level, resource-efficient capabilities of C\texttt{++}. Moreover, in order to delineate the regime of the quantum supremacy/advantage, analogous investigation exhausting classical resources are called for and can be performed in a seamless fashion within a C\texttt{++} framework, while posing many technical obstacles to a high-level language such as Python. With these considerations as a key component governing its design and by delivering a framework that spans the entire quantum computation workflow in a highly extensible fashion, XACC posits itself in the forefront of quantum software for chemical applications, for both the current hybrid quantum-classical regime as well as future fault tolerant machines.

\section{Acknowledgements}

Part of this work was supported by the 'Embedding Quantum Computing into Many-body Frameworks for Strongly Correlated Molecular and Materials Systems' project funded by the U.S. Department of Energy (DOE), Office of Science, Office of Basic Energy Sciences, the Division of Chemical Sciences, Geosciences, and Biosciences. DC and DL would like to acknowledge funding by the US Department of Energy award ERKCG13 provided by the Office of Basic Energy Sciences. AM would like to acknowledge funding by the US Department of Energy Office of Science Advanced Scientific Computing Research (ASCR), Accelerated Research in Quantum Computing (ARQC). This research used resources of the Compute and Data Environment for Science (CADES) at the Oak Ridge National Laboratory, which is supported by the Office of Science of the U.S. Department of Energy under Contract No. DE-AC05-00OR22725. Oak Ridge National Laboratory is managed by UT-Battelle, LLC, for the US Department of Energy under contract no. DE-AC05-00OR22725.

\bibliographystyle{ACM-Reference-Format}
\bibliography{main}

\end{document}